\documentclass{article}
\usepackage[utf8]{inputenc}

\usepackage{epsfig}
\usepackage{graphicx}
\usepackage{dcolumn}
\usepackage{bm}
\usepackage{url}
\usepackage[dvipsnames]{xcolor}
\usepackage[colorinlistoftodos,prependcaption,textsize=small]{todonotes}
\usepackage{natbib}
\usepackage{authblk}
\title{Quantifying dynamical high-order interdependencies from the O-information: an application to neural spiking dynamics}
\author[1,2]{Sebastiano Stramaglia}
\author[1,2]{Tomas Scagliarini}
\author[3]{Bryan C. Daniels}
\author[4]{Daniele Marinazzo}
\affil[1]{Dipartimento Interateneo di Fisica,
Universit\'a degli Studi Aldo Moro, Bari and INFN, Sezione di Bari, via Orabona 4, 70126 Bari, Italy}
\affil[2]{Center of Innovative Technologies for Signal Detection and Processing (TIRES),Universit\'a degli Studi Aldo Moro, Bari, via Orabona 4, 70126
Bari, Italy}
\affil[3]{ASU–SFI Center for Biosocial Complex Systems, Arizona State University, Tempe, Arizona, USA}
\affil[4]{Department of Data Analysis, Ghent University, 2 Henri Dunantlaan, 9000 Ghent, Belgium}

\date{\today}                     
\begin{document}

%
%

\maketitle

\begin{abstract}
We address the problem of efficiently and informatively quantifying how multiplets of variables carry information about the future of the dynamical system they belong to.
In particular we want to identify groups of variables carrying redundant or synergistic information, and track how the size and the composition of these multiplets changes as the collective behavior of the system evolves.
In order to afford a parsimonious expansion of shared information, and at the same time control for lagged interactions and common effect, we develop a dynamical, conditioned version of the O-information, a framework recently proposed to quantify high-order interdependencies via multivariate extension of the mutual information. We thus obtain an expansion of the transfer entropy in which synergistic and redundant effects are separated.
We apply this framework to a dataset of spiking neurons from a monkey performing a perceptual discrimination task. The method identifies synergistic multiplets that include neurons previously categorized as containing little relevant information individually.

\end{abstract}

\section{Introduction}
High-order order interdependencies are at the core of complex systems. In many biological systems, pairwise interactions have been found to be insufficient for explaining the orchestrated activity of multiple components. \citep{CRUTCHFIELD199411,ohiorhenuan2010sparse,katz2011inferring,yu2011higher,daniels2016quantifying}. This may be crucial also in relation with the important question of how organ systems dynamically interact and collectively behave as a network to produce health or disease, the core business of network physiology \citep{NP2012,NP2020}. 
The abundance of available data is pushing nowadays the development of effective algorithms for the inference of higher order interactions from data \citep{PhysRevLett.100.238701,PhysRevE.86.066211}. When an information theoretical point of view is adopted, the problem of higher order interactions becomes related with the decomposition of the information flow in redundant and synergistic components, an issue which cannot be addressed within the Shannon framework unless further assumptions are made \citep{williams2010nonnegative}. Partial Information Decomposition (PID) algorithms have been proposed  \citep{barrett2015exploration,Lizier2018,bertschinger2014quantifying} based on the idea that  synergies are statistical relationships which can be seen only if the whole set of driving variables is considered. Unfortunately, the practical use of PID is greatly limited by the super-exponential growth of terms for large systems, and many works limit the analysis to triplets of variables \citep{marinazzo2019synergy}. 

Transfer entropy \citep{schreiber2000measuring}, which is related to the concept of Granger causality \citep{PhysRevLett.103.238701}, has been proposed to distinguish effectively driving and responding elements and to detect asymmetry in the interaction of subsystems. With the appropriate conditioning of transition probabilities this quantity has been shown to perform better than time delayed mutual information to infer interactions, as delayed correlations often fail to distinguish information that is actually exchanged from shared information due to common history and input signals \citep{bossomaier2016introduction}. The expansion of the transfer entropy from a multiplet of variables to a given target has been developed in \citep{PhysRevE.86.066211} to highlight subgroups of variables which provide redundant and synergistic information to the target. For triplets of variables, the exact calculation of multiscale PID for Gaussian processes has been presented in \citep{Faes2017}.

In a recent paper \citep{PhysRevE.100.032305} a novel quantity has been introduced to study statistical synergy, the O-information, a metric capable of characterising synergy- and redundancy-dominated systems and whose computational complexity scales gracefully with system size, making it suitable for practical data analysis; the O-information has been used to study brain aging in \citep{Gatica2020.03.17.995886}. 
We remark that the O-information uses equal-time samples of variables, so its output depends only on  equal-time correlations in the data-set and is insensitive to dynamic transfer of information; moreover the estimation of O-information does not require a division between predictor and target variables. 

In this work we propose a dynamical generalization of the O-information to handle multivariate time series which, apart from equal-time correlations, takes into account also the lagged correlations with a given variable which is assumed to be the target. The proposed approach highlights informational circuits which dynamically influence the target variable in a synergistic or redundant fashion, with a much lighter computational burden, for large systems,  than those required by the exact expansion of \citep{PhysRevE.86.066211} or PID approaches in the spirit of \citep{williams2010nonnegative}. We apply this quantity, that will be denoted as dO-information, to study the neural spiking dynamics recorded from a multielectrode array with 169 channels during a visual motion direction discrimination task, which has been already considered in \citep{10.3389/fnins.2017.00313} in the frame of Dual Coding Theory; here will analyze this data-set with the aim of characterizing how the dynamic transfer of information is shaped by redundant and synergistic multiplets of variables.

\section{Methods}
Given $n$ random variables arranged in the vector $\mathbf{X}$, the O-information (shorthand for “information about Organisational structure”) is defined as follows \citep{PhysRevE.100.032305}:
\begin{equation}
\Omega_n =(n-2) H(\mathbf{X})-\sum_{j=1}^n \left[ H(X_j)-H(\mathbf{X}\setminus X_j)\right],\label{eq:01}
\end{equation}
where $H$ stands for the entropy. 
If $\Omega_n >0$ the system is redundancy-dominated, while if $\Omega_n < 0$ it is synergy-dominated.
Let us now  add the stochastic variable $y$ to the set of $\mathbf{X}$ variables. The O-information now reads $$\Omega_{n+1}=\Omega_n+\Delta_n,$$ where
\begin{equation}
\Delta_n =(1-n) I(y;\mathbf{X})+\sum_{j=1}^n  I(y;\mathbf{X}\setminus X_j),\label{eq:02}
\end{equation}
$I$ denoting the mutual information. $\Delta_n$ is the variation of the total O-information when the new variable y is added, measuring the informational character of the circuits which link y with variables $\mathbf{X}$: if $\Delta_n$ is positive, then y receives mostly redundant information from $\mathbf{X}$ variables, whilst a negative $\Delta_n$ means that the influence of $\mathbf{X}$ on y is dominated by synergistic effects.

Let us now consider a multivariate set of $n$ time series $\{ x_k\}_{k=1,\ldots,n}$ and a target series $z$. Choosing an order $m$ for the time series, we consider as the random variables $\mathbf{X}$ the state vectors $$X_k (t) =\left(x_k(t)\; x_k(t-1) \;\cdots\; x_k(t-m+1)\right);$$ varying time $t$ we get different samples of $\mathbf{X}$. The role of the variable y is now played by the target time series: $y(t)=z(t+1)$. With these definitions, $\Delta_n$ measures the character of the information flow from the $x$ variables to the target $z$. However, in order to remove shared information due to common history and input signals, one should condition on the state vector of the target variable, thus leading to the definition of the dynamic O-information from the group of variables $\{ x_k\}_{k=1,\ldots,n}$ to the target series $z$:
\begin{equation}
d\Omega_n =(1-n) I(y;\mathbf{X}|Y)+\sum_{j=1}^n  I(y;\mathbf{X}\setminus X_j|Y),\label{eq:03}
\end{equation}
where $Y(t) =\left(z(t)\; z(t-1) \;\cdots\; z(t-m+1)\right).$

We use the expression  of the dO-information  to define the optimization problem of determining the set of k variables which maximizes $d\Omega_k$, with $k < n$; this search leads to the most redundant circuit of k+1 variables, assuming z as the target.
Analogously, the search for the set of k variables which minimizes $d\Omega_k$ leads to the most synergistic  circuit of k+1 variables.  As the extensive search for these motifs is unfeasible for large k, we adopt a greedy search strategy, where the extensive search is performed for $k=2$, and larger k are handled adding one variable at a time to the best multiplet of k-1 variables.

In order to define a criterion to stop the greedy search for the redundant k variables motifs, one can estimate the probability that the increment  $d\Omega_k -d\Omega_{k-1}$ is lower than those corresponding to the inclusion of a randomized time series (obtained, e.g., by  a random circular shift of the k-th selected x time series). The k-th variables is thus added to the multiplet when such probability is lower than a given threshold.

The dO-information is constructed in order to probe higher-order influences, and should be compared to the informational pattern provided by the  information flow network as measured by the pairwise transfer entropy \citep{schreiber2000measuring}: $$TE\left( x_i \to z\right)=I\left( y;X_i|Y\right);$$
to assess the significance of the pairwise transfer entropy, for each pair driver-target we consider surrogate interactions (obtained by blockwise circular shift of the target time series) and accept a non-zero value only if the probability that randomization of the target leads to a value of the transfer entropy higher than the measured one is less than 0.05. Recently another estimator, based on a theoretical
framework for TE in continuous time and extended to event based data, connected to a local permutation surrogate generation strategy, was proposed \citep{shorten2020estimating}.

Further properties of the dO-information are the following. In the case of two driving variables, we have:
\begin{equation}
d\Omega_2 = I(y;X_1|Y)+I(y;X_2|Y)- I(y;X_1 X_2|Y),\label{eq:04}
\end{equation}
coinciding with the second order term of the expansion of the transfer entropy developed in \citep{PhysRevE.86.066211}. Expression (\ref{eq:04}) may be seen as a dynamical generalization of  the interaction information, a well known information measure for sets of three variables \citep{mcgill}.
Another property: let us suppose that the variable $x_n$ is statistically independent of the others, i.e.
\begin{equation}
p\left(y,Y,X_1,X_2,\ldots,X_n\right)=p\left(y,Y,X_1,X_2,\ldots,X_{n-1}\right)\;p\left( X_n\right),\label{eq:05}
\end{equation}

then the dynamic O-information does not change under inclusion of $x_n$, i.e. 
\begin{equation}
d\Omega_n =d\Omega_{n-1}.\label{eq:06}
\end{equation}

Since $d\Omega_1 =0$, the property above ensures that the dO-information is not sensitive to pure pairwise interactions.

\begin{figure}[h!]
\begin{center}
\includegraphics[width=\linewidth]{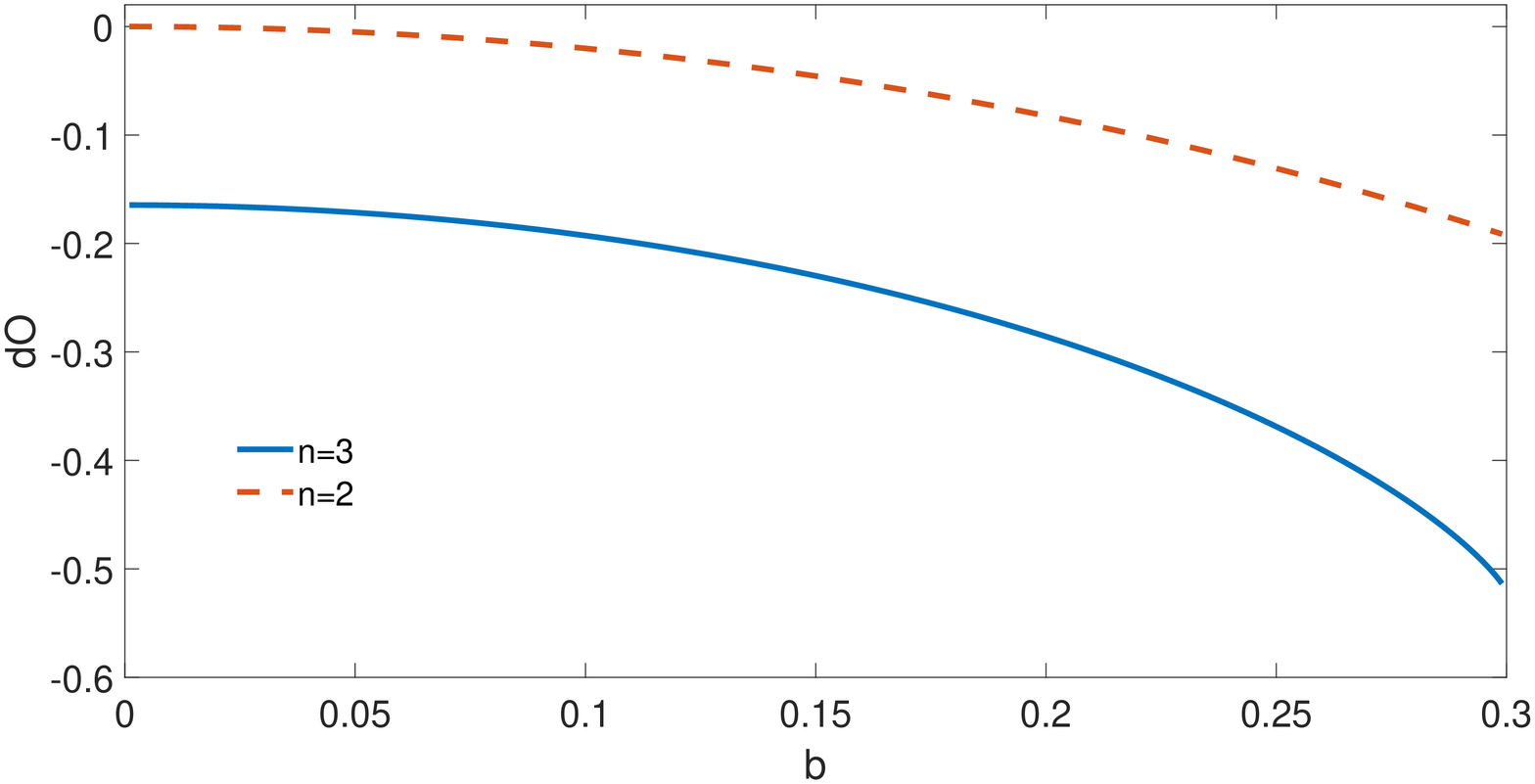}
\end{center}
\caption{For the toy model described in the text, consisting of four binary variables, we depict the dynamic O-Information from the pair of drivers and from the triplet of synergistic drivers as a function of the parameter b; a is fixed at 0.7. }\label{fig:0}
\end{figure}

\begin{figure}[h!]
\begin{center}
\includegraphics[width=\linewidth]{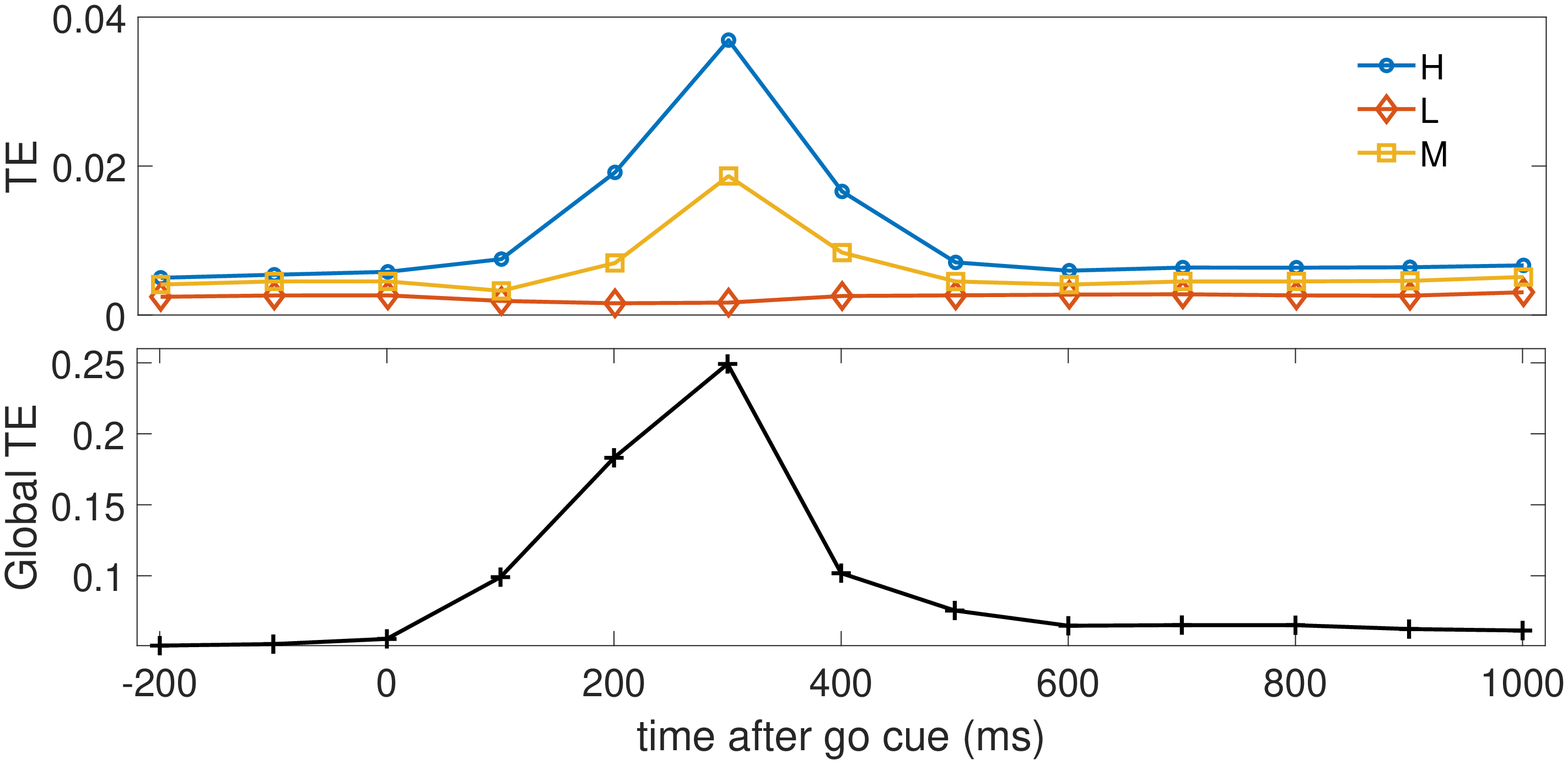}
\end{center}
\caption{Top: the average pairwise transfer entropy towards H neurons is depicted versus time, for three classes of driver: H neurons, M neurons and L neurons.
Bottom: The global transfer entropy is depicted versus time}\label{fig:1}
\end{figure}

We have computed the conditioned mutual information terms, composing the dO-informations, using the Gaussian Copulas approach described in \cite{Ince2017}.

It is worth stressing the conceptual difference between the search for the most informative variables for a given target (i.e. the  n variables $\bf{X}$ maximizing $I(y;\mathbf{X}|Y)$), and the search for the most synergistic multiplet (i.e. the variables $\bf{X}$ minimizing $d\Omega_n$). Suppose that, during the greedy search, one has already selected n-1 variables and now has to look for the n-th variable. If the new variable is selected so as to maximize the information about the target, then the information gain due to its inclusion may be due to synergistic interactions with the previously selected n-1 variables, or to unique information from the new variable, where unique information means a contribution to the predictability of the target that can be obtained from that  variable even when it is treated as the only driver. Inclusion of variables providing unique information does not give us further insights about the system beyond what we already knew from the pairwise description. Minimization of  $d\Omega_n$ is instead tailored to take into account only synergistic interactions, and thus to elicit informational circuits of variables which influence the target: in other words, minimizing $d\Omega_n$ leads to discover (even small) improvements of the predictability of the target which can be ascribed to the joint action of groups of driving variables, thus allowing a picture of the system beyond the pairwise description. 

As a toy example, let consider a system of four binary variables $\sigma_{i}(t)$ such that $\sigma_{1}$ $\sigma_{2}$ and $\sigma_{3}$ are 0 or 1 with equal probability at each time, whilst $P\left(\sigma_{4}(t+1)| \sigma_{1}(t),\sigma_{2}(t),\sigma_{3}(t)\right)$ is given by the following probabilities:
\begin{center}
\begin{tabular}{ ccc } 
$P(0|0,0,0)$&=&1-a+b,\\
$P(0|0,1,0)$&=&a-b,\\
$P(0|1,1,0)$&=&1-a+b,\\
$P(1|1,1,0)$&=&a-b,\\
$P(1|0,1,0)$&=&1-a+b,\\
$P(0|1,0,0)$&=&a-b,\\
$P(1|0,0,0)$&=&a-b,\\
$P(1|1,0,0)$&=&1-a+b,\\
$P(0|0,0,1)$&=&a+b,\\
$P(0|0,1,1)$&=&1-a-b,\\
$P(0|1,1,1)$&=&a+b,\\
$P(1|1,1,1)$&=&1-a-b,\\
$P(1|0,1,1)$&=&a+b,\\
$P(0|1,0,1)$&=&1-a-b,\\
$P(1|0,0,1)$&=&1-a-b,\\
$P(1|1,0,1)$&=&a+b.\\
\end{tabular}
\end{center}

The variable $\sigma_{4}$ receives dynamically synergistic information, by construction, from the pair $\{\sigma_{1},\sigma_{2}\}$ and from the triplet $\{\sigma_{1},\sigma_{2},\sigma_{3}\}$, depending respectively on the parameters b and a.

Indeed, for $b=0$, with probability a the variable  $\sigma_{4}$ at time t+1 is given by the majority rule applied to the three driving variables at time t, unless the three variables are equal: if they are all equal, then  $\sigma_{4}$ becomes the opposite with probability a. Therefore the information provided by  $\{\sigma_{1},\sigma_{2},\sigma_{3}\}$ is synergistic and all the three variables must be known in order to improve the predictability of $\sigma_{4}$. 

On the other hand, for $a=0$, $\sigma_{4}$, with probability b, is given by the XOR applied to $\{\sigma_{1},\sigma_{2}\}$. When both a and b are non vanishing, two synergistic circuits of three and two variables influence the target $\sigma_{4}$.

In figure (\ref{fig:0}) we depict the $d\Omega_3$ from the triplet as well as the $d\Omega_2$ from the pair of synergistic drivers, for a=0.7 and varying b. For b=0, just the triplet $\{\sigma_{1},\sigma_{2},\sigma_{3}\}$ is correctly recognized as driving the target. As b increases, also the pair $\{\sigma_{1},\sigma_{2}\}$ is recognized as a synergistic driver. Note that also $d\Omega_3$ decreases with b: indeed the dynamic O-information $d\Omega_n$ by construction sums up the contributions from the informational circuits corresponding to  subsets of the n variables.

Crucially, in situations like this one, where the information flow is dynamic, the O-information fails to provide a description of the system, and a dynamical approach like 
$d\Omega$ is mandatory.

\section{Dataset}
We use data from the Random Dot Motion discrimination task \citep{shadlen2001neural,kiani2009representation,kiani2014dynamics,kiani2015natural}, in which the subject must decide which direction dots on a screen are moving. This dataset comes from the sample T33 on the data sharing website \url{ https://www.cns.nyu.edu/kianilab/Datasets.html} and has been already analyzed in an information theory framework in \citep{10.3389/fnins.2017.00313}. We
analyze the activity of 169 neural channels in a macaque monkey performing the task, across 1778 trials.  In each trial, after the perceptual stimulus, a go cue is given to the subject to prompt it to indicate its decision. In \citep{10.3389/fnins.2017.00313}  the analysis had decision as the target, and neurons were divided in three groups according to the information that their dynamics provide about the decision: those in Class H encode information before the go cue, those in Class M encode information after but not before the go cue, and those in Class L never encode information.
Here we will not take into account the decision as a variable, retaining only the classification of neurons in the three classes H, M and L.

\begin{figure}[h!]
\begin{center}
\includegraphics[width=\linewidth]{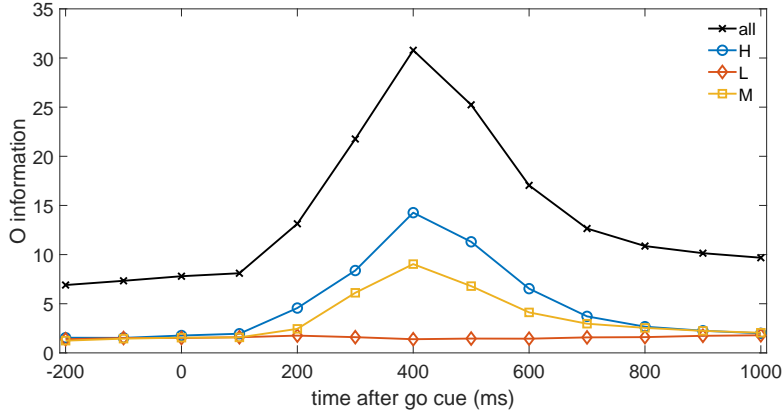}
\end{center}
\caption{The O-information is depicted versus time for the three systems of neurons H, M and L, as well as for the whole system of neurons. }\label{fig:statica}
\end{figure}

As an example application of the proposed method, we will consider the internal dynamics of the neuronal system. For each H neuron, taken as the target, we will study the higher order interactions from the rest of the measured neurons, concentrating on the most redundant circuits as well as the most synergistic ones. 

\begin{figure}[h!]
\begin{center}
\includegraphics[width=\linewidth]{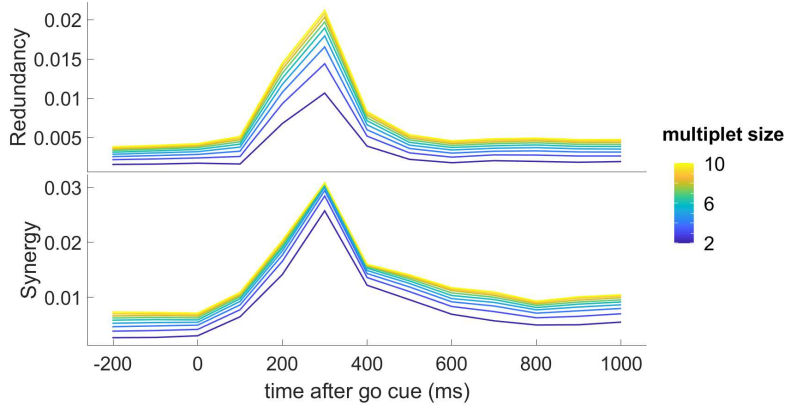}
\end{center}
\caption{Top: the redundancy ($d\Omega_k$) from the optimal k-multiplet to an H neuron, as found by greedy search, is depicted as a function of time for k ranging from 2 to 10; the plotted quantity is the average over all the H target neurons. Bottom: the synergy ($-d\Omega_k$) from the optimal k-multiplet to an H neuron, as found by greedy search, is depicted as a function of time for k ranging from 2 to 10; the plotted quantity is the average over all the H target neurons.}\label{fig:2}
\end{figure}

\section{Results}
For the following analyses we used $m=1$ as lag for conditioning in the past, namely a bin one time step in the past. In figure (\ref{fig:1}) we depict the pairwise transfer entropy as a function of time, where the target is an H neuron and the driver is a neuron belonging to one of the three classes; the curves are averaged over the target neuron and over the driver belonging to each of the three classes. We observe that the information flow peaks around 300ms after the go cue, and that most of the effective influence arise from the other H neurons and (to a lesser extent) from M neurons. The pairwise transfer entropy from L neurons is negligible, hence at the bivariate level L neurons seem to play no role in the construction of the dynamical response of the system. The lower panel of the figure depicts the global transfer entropy \citep{barnett2013information} averaged over all the H neurons as targets; the global transfer entropy measures the information flow about the target when all the other variables are simultaneously taken as the driving set.   The global transfer entropy of a kinetic Ising model has been shown to have a maximum in the disordered phase \citep{barnett2013information}. Successively it has been shown \citep{marinazzo2019synergy} that it is the synergistic component of it that is responsible for this peak, which can be considered as an early warning of a transition towards order.

\begin{figure}[h!]
\begin{center}
\includegraphics[width=\linewidth]{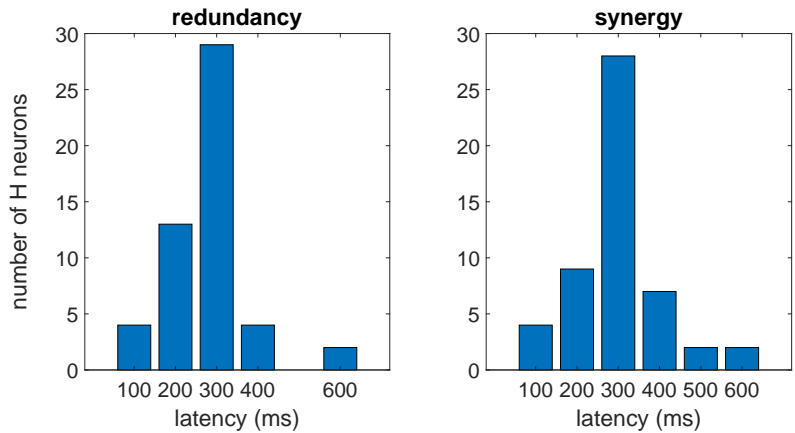}
\end{center}
\caption{Each H neuron experiences a peak of the redundancy (synergy) whose latency can vary from neuron to neuron. In this figure we consider the optimal multiplet of 10 variables, plotting the distribution of latencies both for the redundant (left) and synergistic one (right).}\label{fig:3}
\end{figure}

Let us now turn to consider higher order interactions, and start with the O-information, the approach introduced in \cite{PhysRevE.100.032305} which considers only equal time correlations. In figure (\ref{fig:statica}) we depict, as a function of time, the O-information of the three sets of neurons as well as the O-information of the whole system of neurons. We note that H neurons (and M neurons as well) increase their redundancy (as measured by O-information) with a latency of 400 ms, where also the whole system of neurons displays a clear peak. On the other hand the system of L neurons do not show any reaction to the go cue at the level of O-information. 

To take into account dynamic transfer of information, we apply the proposed approach:
for each target H neuron we look for the multiplet of k variables maximizing $d\Omega_k$ and we call {\it redundancy} the value of the maximum, as described in the Methods Section. Analogously we look for the multiplet of k variables minimizing $d\Omega_k$ and we call {\it synergy} the opposite of the value of the minimum. In figure (\ref{fig:2}) we depict the redundancy and the synergy, as a function of time and for k ranging from 2 to 10; this figure shows that the response to the stimulus is also shaped by higher-order influences, both of the redundant and synergistic types. Also higher-order infleunces peak at 300ms, and the synergistic influences seem to show a slower decay after the peak. 

It is also interesting to note that the incoming synergy and redundancy peak with a latency which slightly depends on the target neuron.  In figure (\ref{fig:3}) we plot the distributions of latencies of maximal redundancy and synergy in optimal multiplets of 10 variables, suggesting that the synergistic response occurs, on average, slightly later than the redundant response.

\begin{figure}[h!]
\begin{center}
\includegraphics[width=\linewidth]{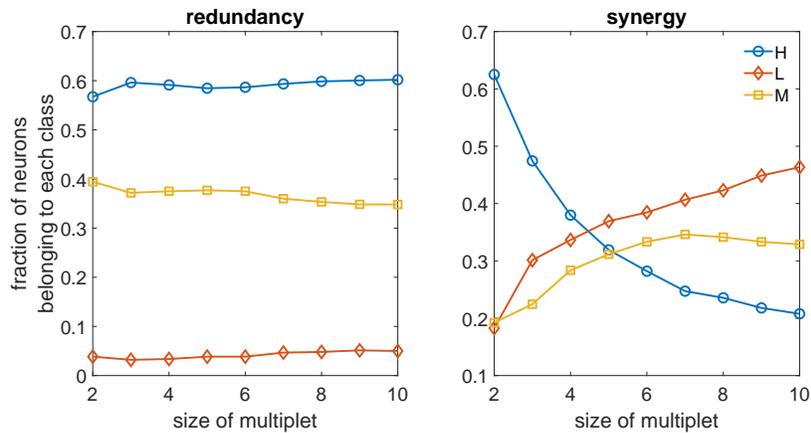}
\end{center}
\caption{Left: as a function of the size of the optimal multiplet, we depict the typical fraction of H, M, and L neurons constituting the multiplet, both for redundancy (left) and synergy (right)}\label{fig:4}
\end{figure}

In figure (\ref{fig:4}) we depict, as a function of the number of driving variables k, the fraction of variables in the best redundant multiplet belonging to the three classes. We observe that redundant circuits are made of H and M neurons, L neurons rarely appearing in the redundant circuits. On the other hand, we see that L neurons can play a relevant role in synergistic circuits as k becomes larger, and are more important than H and M neurons in the construction of synergistic circuits.

\begin{figure}[h!]
\begin{center}
\includegraphics[width=\linewidth]{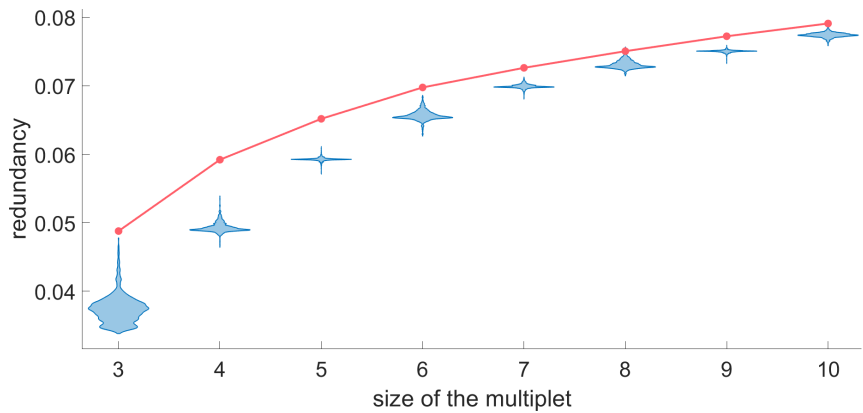}
\end{center}
\caption{For a representative H neuron, the red line represents the redundancy as a function of the size of the optimal multiplet.Each violin plot represents 30.000 realizations of $d\Omega_k$ obtained by a random circular shift of the k-th variable of the multiplet. We accept as truly redundant the multiplets with significance of $5\%$ after Bonferroni correction. Since the null hypothesis cannot be rejected at $k=8$, we conclude that a redundant circuit of 7 driving variables exists influencing the given H neuron.}\label{fig:5}
\end{figure}

In figure (\ref{fig:5}) we show how one can statistically assess the significance of a redundant or synergistic circuit linked to a target, choosing as the target a randomly selected H neuron. While adding a variable to the redundant multiplet with the greedy search, we also evaluate the redundancy that would be obtained using a randomized version of that variable, and we accept that variable if the probability to get an higher value of the redundancy, after randomization, is less than 0.05 after Bonferroni correction. For the target neuron under consideration we find that the multiplet with 7 driving variables can be considered statistically significant, as the null hypothesis can be rejected for $k \leq 8$.

\begin{figure}[h!]
\begin{center}
\includegraphics[width=\linewidth]{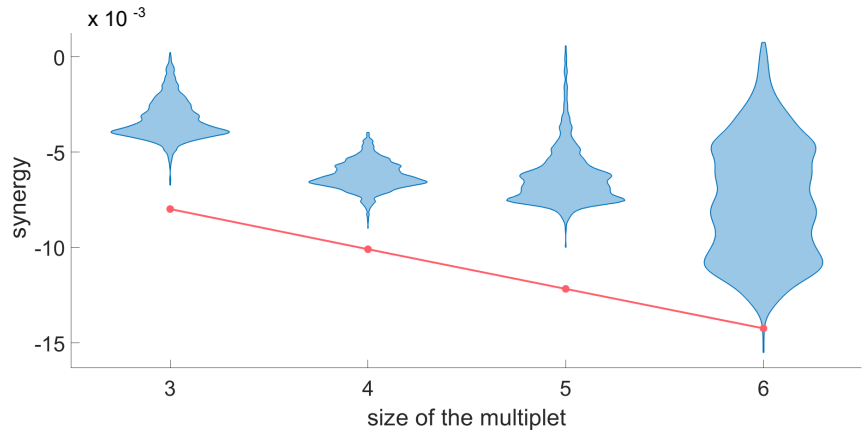}
\end{center}
\caption{ For a typical H neuron, the red line represents $d\Omega_k$ in the synergistic search,  as a function of the size of the multiplet. Each violin plot represents 30.000 realizations of $d\Omega_k$ obtained by a random circular shift of the k-th variable of the multiplet. We accept as truly synergistic the multiplets with significance of $5\%$ after Bonferroni correction. Since the null hypothesis cannot be rejected at $k=6$, we conclude that a synergistic circuit of 5 driving variables exists influencing the given H neuron.}\label{fig:6}
\end{figure}

In figure  (\ref{fig:6}) we do the same for the synergy using the same target neuron. Since the null hypothesis cannot be rejected at $k$ equal to six, we conclude that the synergistic circuit of five driving variables influencing the target is the largest synergistic multiplet that we can assess statistically.

Some of the figures were generated with Gramm \citep{morel2018gramm}.

\section{Discussion}
We have proposed a novel approach to analyze higher-order dynamical influences in multivariate time series, and to highlight redundant and synergistic groups of variables influencing a given target variable. Our method generalizes to the dynamic case a recently introduced quantity, named O-information, which was proposed to assess the informational character of equal-time correlations in a set of random variables \citep{PhysRevE.100.032305}. Our conditioned approach has the main advantage of allowing the distinction of information that is actually exchanged from shared information due to common history and
input signals. Compared with the expansion in \citep{PhysRevE.86.066211} or PID decompositions in the spirit of \citep{williams2010nonnegative}, the proposed approach is computationally much more feasible. However our approach focuses only on finding multiplets that are synergy-dominated or redundancy-dominated, and the corresponding values of synergy and redundancy do not come from  an exact decomposition of the information flow. For this reason their magnitudes cannot be easily compared for varying $k$, but in our opinion this is a reasonable price to pay in order to have a fast algorithm that can handle big data sets.

We believe that our approach can have wide applicability in physiology, in particular at the system level where higher-order interactions may play a role in the collective regulation of dynamical rhythms in the human body \citep{NP2015}.

The relation between mutual information and synergistic information processing in spiking neurons from organotypic cultures of mouse neocortex was recently addressed in \citep{sherrill2020correlated}, and was found to depend on the timescale and the degree of correlation in neuronal interactions. As an example of application of dO-information, we have considered the response of a neural system to an external stimulus. We have shown that, in addition to higher order equal time interactions,  which show a peak for the redundancy (as probed by the O-information) 400 ms after the go cue, the system displays also significant dynamic transfer of information consisting in synergistic and redundant circuits peaking 300 ms after the go cue. 
A recent study on computing TE between spiking neurons \citep{shorten2020estimating} presented some results on the dependency of the values of TE on the firing rate. Based on these estimations, and given the number target events in the present experiment we can expect that the height of the peak of the TE in figure 2 could be slightly overestimated, given the increased firing rate in the same interval. On the other hand the bias is stronger, and towards positive values, with a reduced number of spikes, and the low values before and after the peak are an indication that the TE peak itself is meaningful. The results are further backed up by the surrogate procedure.

Concerning the dynamics of H neurons,from the point of view of pairwise influence, H neurons are the most important drivers, M neurons are also relevant but to a lesser extent, whilst L neurons do not play any role. Going beyond the pairwise description, as far as the redundancy is concerned we find the same relative contributions in terms of the composition of redundant multiplets: the abundance of H neuron is higher than those of M neurons whilst the contribution of L neurons is negligible. On the other hand, considering the synergy the relative importance of the three types of neurons is changed: for large multiplets the abundance of L and M neurons is higher than those of H neurons, thus suggesting that surprisingly also L neurons may play a role in shaping the dynamics of H neurons by participating in synergistic groups of variables. We have shown that synergy of multiplets of variables can take values up to 0.03 bits. It is worth stressing that dO is not derived from an exact decomposition of transfer entropy and that this value cannot be interpreted as a gain in predictability of the target; however it suggests that the role played by synergistic circuits is small but not negligible when compared with 0.25 bits which is the peak of the global transfer entropy to H neurons, when all the other neurons are simultaneously taken as the driving set. Further investigations are certainly needed to confirm the role of M and L neurons in the higher order description of dynamics of H neurons; our analysis shows that the proposed approach is capable of highlighting these effects while requiring a reasonable computational effort.

\section*{Funding}
SS and TS acknowledge support of the Italian Ministry for University and Research (MIUR), project PRIN 2017/WZFTZP "Stochastic forecasting in complex systems" 

\section*{Acknowledgments}
The authors thank Bill Newsome and Roozbeh Kiani for use of the data.

\bibliography{test}

\end{document}